\documentclass[structabstract]{aa}  
\usepackage{graphicx}
\begin{document}

\title{Comments on ``The long-period Galactic Cepheid RS Puppis. I. A
geometric distance from its light echoes''}


\author{Howard E. Bond
\and
William B. Sparks
}

\institute{Space Telescope Science Institute, 3700 San Martin Dr., Baltimore,
MD 21218, USA\\
\email{bond@stsci.edu, sparks@stsci.edu}
}

\date{Received \quad \quad \quad ; accepted \quad \quad \quad }

 
\abstract
{The luminous Galactic Cepheid RS~Puppis is unique in being surrounded by a dust
nebula illuminated by the variable light of the Cepheid. In a recent paper in
this journal, Kervella et al.\ (2008) report a very precise geometric distance
to RS~Pup, based on measured phase lags of the light variations of individual
knots in the reflection nebula.}
{In this commentary, we examine the validity of the distance measurement, as
well as the reality of the spatial structure of the nebula determined by Feast
(2008) based upon the phase lags of the knots. }
{Kervella et al.\ assumed that the illuminated dust knots lie, on average, in
the plane of the sky (otherwise it is not possible to derive a geometric
distance from direct imaging of light echoes).  We consider the biasing
introduced by the high efficiency of forward scattering.}
{We conclude that most of the knots are in fact likely to lie in front of the
plane of the sky, thus invalidating the Kervella et al.\ result. We also show
that the flat equatorial disk structure determined by Feast is unlikely;
instead, the morphology of the nebula is more probably bipolar, with a
significant tilt of its axis with respect to the plane of the sky.}
{Although the Kervella et al.\ distance result is invalidated, we show that
high-resolution polarimetric imaging has the potential to yield a valid 
geometric distance to this important Cepheid.}

\keywords
{stars: individual: RS Pup --
stars: circumstellar matter --
stars: distances --
stars: variables: Cepheids --
ISM: reflection nebulae --
scattering}

\authorrunning{Bond \& Sparks}
\titlerunning{Critique of RS Puppis Distance}

\maketitle


\section{Introduction}

\def\Afsar{Af\c{s}ar}
\def\HST{{\it HST}}
\def\RS{RS~Pup}

RS~Puppis (pulsation period 41.4~days) is one of the most luminous known
Galactic Cepheids. It is thus a Milky Way analog of the bright Cepheids used to
determine extragalactic distances, calibrate the brightnesses of Type~Ia
supernovae, and determine the Hubble constant. An accurate distance to \RS\
would provide a valuable anchor in setting the zero-point for the extragalactic
distance scale and in determining whether there is a change in slope for the
Cepheid period-luminosity relation (PLR) at long periods, as some authors have
discussed (e.g., Ngeow et al.\ 2008 and references therein). Unfortunately, \RS\
is too far away for a trigonometric parallax measurement with existing
instrumentation, even with the Fine Guidance Sensors onboard the {\it Hubble
Space Telescope\/} (\HST\/) (Benedict et al.\ 2007).

\RS\ is nearly unique among Galactic Cepheids in being embedded in a reflection
nebula, as first pointed out by Westerlund (1961). Havlen (1972) demonstrated
that features within the nebula show light variations at the Cepheid's pulsation
period and argued that these ``echoes'' could be used to derive a geometric
distance. More recently,  Kervella et al.\ (2008, hereafter K08) have used the
3.6-m ESO NTT telescope to monitor light variations within the nebula in
excellent seeing. K08 were able to identify nearly three dozen nebular knots
showing variations at the Cepheid period, of which the 10 best features were
selected for analysis. By measuring the angular separations from the star, and
determining the phase lags for each of these knots with respect to the stellar
pulsation, they determined a geometric distance to \RS\ of $1992\pm28$~pc. If
valid, this would be by far the most precise distance measured for any Cepheid,
Galactic or extragalactic (in terms of percent error), and it would provide a
high-weight zero-point for the PLR, extragalactic distances, and $H_0$. It is
important to examine the validity of this result, especially because it is about
15\% higher than the 1728~pc found for \RS\ by Feast (2008) through application
of the PLR of van~Leeuwen et al.\ (2007); these latter authors have recently
re-calibrated the zero-point on the basis of \HST\/ and revised {\it
Hipparcos\/} trigonometric parallaxes of nearby Cepheids (none of which,
however, have periods as long as that of \RS).

Unfortunately, in spite of the spectacular quality of their observations, we
will demonstrate that K08's analysis is based on an unjustified assumption that
invalidates their result. We will argue that a geometric distance cannot be
obtained from any direct imaging of the \RS\ nebula. However, we will also show
that the addition of polarimetric imaging has the potential to enable a truly
direct geometric distance determination for this important object.

\section{Light-Echo Geometry}

The geometry of a light echo is simple, in the case of a single flash from the
illuminating star. At a time $t$ since outburst, the illuminated dust lies on
the paraboloid given by $z=x^2/2ct-ct/2$, where $x$ is the projected separation
from the star in the plane of the sky, and $z$ is the distance along the line of
sight toward Earth. Thus the outer parts of the echo usually correspond to dust
{\it in front of\/} the star, and the inner parts to dust {\it behind\/} the
star. Because of the geometry, the outer parts of a light echo generally appear
to expand at a speed much greater than the speed of light.

The most spectacular light echo in astronomical history is on display around the
exotic variable star V838~Monocerotis (Bond et al.\ 2003; Corradi \& Munari
2007), imaged extensively by \HST\/ (Bond 2007). Direct imaging of the echo
cannot yield a geometric distance. However, {\it polarimetric imaging\/} can
identify the region within the echo where the linear polarization is maximum,
marking the location where the scattering angle toward the Earth is $90^\circ$.
This highly polarized light corresponds to dust lying {\it in the plane of the
sky}, where the light-echo equation shows that the linear distance from the star
is $x=ct$. The angular separation of this location from the star then yields the
geometric distance. This novel method for astronomical distance determination
was first spelled out by Sparks (1994). Sparks et al.\ (2008, hereafter S08)
have now applied the method to \HST\/ polarimetric images of V838~Mon obtained
in 2002--5. They indeed detected a ring of highly polarized light expanding at
the speed of light, and found a distance of $6.1\pm0.6$~kpc. This determination
has been verified independently through classical main-sequence fitting for a
sparse stellar cluster associated with V838~Mon by \Afsar\ \& Bond (2007),
yielding a distance of $6.2\pm1.2$~kpc.

\section{The Nested Light Echoes of RS Puppis}

The light-echo situation is far more complicated when the illuminating source is
a periodic variable star. In this case, as seen from Earth, a train of nested
paraboloids propagates outward into the surrounding dust. This geometry is
illustrated in Fig.~1. The illumination seen at a given projected separation,
$x$, from the star will be the integral along the line of sight of the incident
intensity at each $z$ location (a function of the corresponding phase of the
Cepheid variation), modified by the dust density at each volume element, by
$r^{-2}$, where $r$ is the distance of that volume element from the star, and by
the dust scattering phase function $\Phi(\xi)$, where $\xi$ is the scattering
angle (taken to be $0^\circ$ for forward scattering). See S08 for a detailed
discussion of such calculations and references to earlier literature. 

Fig.~1 thus demonstrates that, for a uniformly filled nebula, the light
variations will largely average out over the line of sight, since all phases of
the light variation are covered multiple times. {\it Only if there is a
non-uniform distribution of dust along the line of sight (for example, isolated
nebular knots, or a very non-spherical nebular morphology) will we see
high-amplitude periodic light variations at those locations.} K08 in fact did
confirm the work of Havlen by demonstrating the existence of locations in the
nebula where there are such variations.

However, even for such isolated knots, there is an unavoidable ambiguity in
their $z$ locations. Referring again to Fig.~1, let us consider the brightest
variable knot found by K08, which is ``Knot~5'' in their Table~3. It lies at a
separation of $x=16\farcs03$ from the star, marked as a vertical dashed straight
line in Fig.~1. The fractional phase lag of Knot~5 is observed to be 0.504,
i.e., its light maxima lag about 20.9~days behind those of \RS\ itself. Thus its
$z$ location must lie halfway between a pair of parabolas along the straight
line in Fig.~1, {\it but we do not know which pair!}  The knot could lie
$\sim$0.23~pc in front of the plane of the sky (phase lag 1.504 cycles), or
0.10~pc (2.504 cycles), or 0.02~pc (3.504 cycles), or in the plane of the sky
(4.504 cycles), or 0.02~pc behind (5.504 cycles), etc.  (The $z$ distances given
here depend, of course, on the assumed stellar distance, in this case 2~kpc.)
When the only observations available are direct images, {\it there is no way to
resolve this uncertainty without making arbitrary assumptions.}

K08 resolved the ambiguity by {\it assuming\/} the knots to lie in the plane of
the sky. Under this assumption, and making an initial guess at the distance to
the star, the integer numbers of cycles in the phase lags become known (in the
case of Knot~5, it is 4). Then the fractional phase lag for each knot yields a
stellar distance, and the final result was taken as the mean of the distances
determined from 10 individual features, i.e., $1992\pm28$~pc. 

K08 recognized that the assumption that the knots lie in the plane of the sky
would generally not be true for any particular individual knot, but argued that
the assumption would be true in the mean, as long as the spatial distribution of
the knots is close to isotropic.\footnote{Havlen (1972, and 2008 private
communication) made an additional argument that
the variable knots appear to be limb-brightened edges of shells centered on
RS~Pup, and noted that to the extent these shells are spherical and centered on
the star, they would tend to lie in the plane of the sky. Unfortunately, this
argument is somewhat self-contradictory, since limb-brightening requires a
significant physical extent along the line of sight, which would tend to average
out the light variations.}

\section{Critiques of the K08 Analysis}

\subsection{Summary of K08 Analysis Procedure}

The \RS\ nebula was imaged by K08 using the 3.6-m European Southern Observatory
New Technology Telescope (NTT) and the ESO Multi-Mode Instrument (EMMI)\null.
CCD images were obtained on 7 nights between 2006 October and 2007 March, in
seeing generally ranging from $0\farcs8$ to $1\farcs1$.

The analysis procedure employed by K08, based on their superb time series of
high-resolution images of the nebula, is first to isolate a set of locations
within the nebula where there are light variations at the Cepheid's pulsation
period, and where the amplitudes of the variations are large and comparable to
that of the Cepheid itself. The large amplitude ensures that the knot is
isolated and that additional material elsewhere along the line of sight, and
thus out of phase, is not significantly diluting the variations. K08 identified
10 prominent features that satisfied these criteria, as well as a larger number
of lower-weight knots.

At each such location in the nebula, there are two observables: the angular
separation of the knot from the star, and the phase lag of its light curve
relative to RS~Pup. In the notation of K08, the angular separation of knot $i$
is $\theta_i$, and its phase lag consists of an integer number of
periods, $N_i$ (which is unknown), plus a fractional lag, $\Delta\phi_i$,
determined from the amount of time by which the echo light curve is out of phase
with the Cepheid.

With these two observables, the K08 procedure is to assume that the knot lies in
the plane of the sky, adopt an initial estimate of the distance $d$, and then
derive the integer number of periods in the time delay, according to $N_i={\rm
int}(\theta_i \, d/cP)$, where $c$ is the speed of light and $P$ is the
pulsation period of the Cepheid. Once $N_i$ is known for each knot, a distance,
$d_i$, is obtained based on each one, using $d_i =
(N_i+\Delta\phi_i)P/c\theta_i$. If $\theta_i$ is in arcseconds, and $P$ in days,
this equation becomes $d_i = (N_i+\Delta\phi_i)P/(5.7755\times10^{-3} \,
\theta_i)\, \rm pc$. K08 adopt a period of $P=41.4389$~days. 

Using these equations, K08 stepped through a range of initial estimates of $d$,
calculated $N_i$ and the corresponding $d_i$ for each knot, and then derived the
standard deviation of the $d_i$ values. They found a deep minimum of the
dispersion of the distance estimates at a weighted value of $1992\pm 28$~pc, as
discussed in detail in their paper.


\subsection{Statistical Significance}

We first present a simple test of the statistical significance of the above
result, which was based on searching for the minimum standard deviation of the
distance estimates.  We follow the procedure of K08, namely stepping through a
range of starting values of the distance, and calculating the relative standard
deviation (the standard deviation divided by the mean distance) at each assumed
distance. We use the set of 10 high-weight pairs of $\theta_i , \Delta\phi_i$
values given by K08 (their Table~3), and follow their procedure of deleting the
two lowest and two highest values of $d_i$ before calculating the mean and
standard deviation.  The top left panel in Fig.~2 plots the relative standard
deviation vs.\ assumed distance for the K08 data. This plot shows a minimum
relative standard deviation of 1.6\% at a distance of 1975~pc (K08 give 1.1\% at
1992~pc, due presumably to a slightly different weighting and/or deletion
scheme).

To test the significance of this result, we tried the experiment of adopting the
same set of 10 values of $\theta_i$, but assigned phase lags, $\Delta\phi_i$,
generated by a random-number routine that outputs values in the 0.00--1.00
range. We ran this experiment 9 times, and plot the results in the remaining 9
panels in Fig.~2.  As the panels show, two-thirds of these randomizations (6
out of 9) show {\it lower\/} minima than the 1.6\% obtained for the actual data.
(The lowest minimum found in the range 0 to 3000~pc was a relative standard
deviation of only 0.6\% in the 8th randomization, at a distance of 2380~pc.)

We believe that this simple experiment demonstrates very low statistical
significance for the distance of 1992~pc found by K08 through minimizing the
relative standard deviation of the distance estimates. Note that the relative
standard deviation systematically trends downward with increasing assumed
distance. This arises naturally because the cycle count, $N_i$, increases with
assumed distance, but the phase lags can never vary by more than $\pm$0.5; thus
the percentage scatter in the derived distances decreases systematically with
distance.


%

\subsection{Biasing Away from the Plane of the Sky}

However, there is a more fundamental objection to the K08 analysis.
Unfortunately, the assumption that the light-variable knots are in the plane of
the sky is quite unlikely to be true, not even in the mean, and not even if the
knots are distributed isotropically around the star. In fact, we expect a strong
biasing such that the brighter knots are likely to lie {\it in front of\/} the
plane of the sky, because of the behavior of the scattering phase function
$\Phi(\xi)$. 


This phase function is usually taken to be that of Henyey \& Greenstein (1941):
$\Phi(\xi) = (1-g^2)/[4\pi (1+g^2-2g\cos\xi)^{3/2}]$, in which the parameter $g$
for interstellar dust is empirically found to be near $g=0.6$ (see references
in S08). In this case, for example, the relative scattering efficiencies for
angles of $0^\circ$, $15^\circ$, $30^\circ$, $45^\circ$, $90^\circ$, and
$180^\circ$ are 1, 0.71, 0.35, 0.17, 0.04, and 0.016. Thus knots with smaller
scattering angles will be systematically brighter, and as Fig.~1 shows these
knots {\it are preferentially in the foreground.} Hence the high efficiency of
forward scattering imposes an unavoidable biasing away from knots lying in the
plane of the sky.  

A quantitative illustration of this biasing is shown in Fig.~3. Here we have
taken Fig.~1 and superposed regions that are color-coded according to the local
values of $\Phi(\xi)/r^2$, where $\Phi(\xi)$ is the dust scattering
efficiency described above, with $g=0.6$, and $r$ is the distance of the volume
element from the star. Thus this function, multiplied by the local volume
density of dust, is proportional to the observed surface brightness at that
location in the nebula as seen from Earth.\footnote{We are assuming that the
dust is sufficiently optically thin that multiple scattering is negligible. A
high optical depth would, of course, provide yet another source of biasing
toward the foreground.} The contours between the colored regions correspond to
steps of factors of 2 in $\Phi(\xi)/r^2$. To interpret the figure, consider
for example a dust knot located at a projected separation of $x\simeq0.15$~pc
from the star, typical of the variable features found by K08. The same knot, if
located $\sim$0.15~pc in front of the plane of the sky would be about 2 times
brighter than if it were located in the plane, and about 9 times brighter than
if it were located $\sim$0.15~pc behind the plane.  At $x\simeq0.1$~pc, a knot
located $\sim$0.25~pc in front of the plane of the sky would be 1.8 times
brighter than if in the plane of the sky, and 32 times brighter than the same
knot located 0.25~pc behind the plane.  More generally, there are large extents
along the positive $z$ direction of nearly constant brightness, but once the
plane of the sky is crossed the brightness drops very rapidly.\footnote{Fig.~3
of course also explains why the surface brightness of the light echo of V838~Mon
faded so rapidly with time, as its single echo paraboloid propagated to the side
and into the background.}

%

There is also a second and more subtle effect that likewise biases the variable
knots toward lying in front of the plane of the sky. For a knot to show a
large-amplitude light variation, it must be spatially isolated so as not to have
its variation diluted by scattering from other material in the same line of
sight, illuminated by light with a range of different phase lags. As Figs.~1 and
3 illustrate, the echo paraboloids become more closely spaced in the
$z$-direction as we move down the line of sight. This means that a knot with the
same linear extent along the line of sight will suffer increasingly more phase
dilution as it is placed at successively greater distances from the observer
along the $z$-axis.

Because of this strong biasing of the illuminated and high-amplitude knots to
lie in the foreground, the assumption of K08 that they lie in the plane of the
sky appears unjustified.  {\it When this assumption has to be abandoned, it is
no longer possible to derive a geometric distance from the phase lags.}


The assumption that a knot lies in the plane of the sky when it in fact does not
will lead to very large errors in the derived distances. For a concrete example,
we consider again Knot~5, lying at $\theta_5 = 16\farcs03$, which was discussed
in \S3. If this knot is assumed to lie in the plane of the sky, the distance
will be calculated to be 673, 1120, 1568, and 2015~pc, for adopted values of
$N_5$ of 1, 2, 3, and 4, respectively. K08 had assumed $N_5=4$. Actually, as
shown in Fig.~3, Knot~5 is statistically likely to lie $\sim$0.2~pc in front of
the plane of the sky and to have $N_5\simeq1$ (if the distance to the star is of
order 2~kpc).

\section{Three-Dimensional Structure of the RS Puppis Nebula}

\subsection{Equatorial Disk?}

In a paper that came to our attention as we were completing this critique,
Feast (2008) discusses the results of K08 and reaches the same conclusion that
we do, namely that the assumption that the echo knots lie in the plane of the
sky is unjustified. 

Feast then essentially reverses the analysis of K08, using the measured phase
lags in order to study the three-dimensional distribution of the light-variable
nebular knots. He {\it assumes\/} a distance to \RS\ of 1728~pc based on the
PLR, and then uses the $\Delta\phi_i$ values to solve for the $z$ locations of
each of the knots. If the knots are not in the plane of the sky, the equation in
\S4.1 becomes $(N_i+\Delta\phi_i) = 5.7755\times10^{-3} D \, \theta_i
(1+\sin\alpha_i)/P\cos\alpha_i$, where $D$ is the distance to the star in pc,
and the angle $\alpha_i$ is, in our notation, given by $\alpha_i=-\tan^{-1}
z_i/x_i$. This equation can be solved to obtain a $z$ depth for each illuminated
knot. Of course, a unique solution is not possible, as the $z$ location depends
on the arbitrarily chosen integer number of cycles in the phase lag, $N_i$.
Feast was, however, able to find a particular set of values of $N_i$, slightly
modified from those obtained by K08 who assumed the knots to lie in the plane of
the sky, which places the knots on the north side of the star in front of, and
those on the south side behind, the plane of the sky. In his solution, the knots
lie close to an inclined plane, which he interprets as indicating that the
nebula is a highly flattened disk, tilted by only $8\fdg1$ to the plane of
the sky. In the top panel of Fig.~4 we have reproduced Feast's Fig.~1, except
that we use the same scale for both coordinates so as to show the correct
proportions. Feast's calculations were made for an expanded set of data for 31
knots, provided by Kervella to Feast and reproduced in the latter's Table~1. In
Feast's notation, $d_y$ is our $-z$, and $d_x$ is the projected separation of
the knot from the line where the plane of the disk intersects the plane of the
sky, given by $d_x = 0.008377 \, \theta_i \sin(\beta_i-\gamma)$~pc, where
$\beta_i$ is the position angle (PA) of the knot, and $\gamma$ is the azimuth at
which the plane of the disk cuts the plane of the sky, taken by Feast to be
$80^\circ$.

Feast's result is surprising because most of the knots are found to be behind
the plane of the sky, contrary to the expectation established here that they
should preferentially lie in the foreground. However, Feast's conclusions are
entirely dependent upon an arbitrary assumption of values of $N_i$ that are
large on the north side, and smaller on the south side, in order to yield this
tilted geometry. To illustrate the arbitrary nature of this analysis, we
repeated his calculations with our own set of $N_i$ values. By visual inspection
of the K08 image of the nebula, it appears to us that there is a long axis
oriented at a PA of about $35^\circ$, which would imply\footnote{We implicitly
assume here that the nebula approximates an elongated structure whose axis has a
PA of about $35^\circ$, rather than a circular disk whose intersection with the
plane of the sky would have a PA of $35^\circ$.} $\gamma=125^\circ$. With this
new value, and by a process of trial and error, we were easily able to obtain a
new set of $N_i$ values that gives the plane containing the knots the {\it
opposite\/} tilt and nearly the same level of scatter, as Feast's fit. Our plot
is shown in the middle panel of Fig.~4.

This exercise primarily serves to demonstrate that there is no unique
three-dimensional structure that can be inferred from the set of
$\theta,\Delta\phi$ pairs.  Our result does have the advantage that the knots on
the south side of the nebula, where they are brightest (see below), are well in
front of the plane of the sky. However, as we argue in the next subsection, it
is more likely that the nebula actually has a pronounced bipolar morphology.

\subsection{Bipolar Nebula?}


We can gain a general idea of the overall morphology and orientation with
respect to the plane of the sky of the \RS\ nebula through ratio imaging. As a
preliminary example of how such a study might proceed, we present in Fig.~5 two
images of the nebula, obtained by W.B.S. with the 3.6-m ESO NTT and EFOSC CCD
coronagraphic imager on 1995 March~5 and May~4. The time interval between the
images was 59.86~days, or 1.44~cycles of the Cepheid. 

The first two frames in Fig.~5 show the two direct images. They were obtained
with an $R$ filter combined with polarizers at four different electric-vector
angles, and these images are the sums of the four. The third frame in Fig.~5
shows the ratio of the May and March images,\footnote{The ratio image and one of
the direct images, but at different stretches from those here, have been
presented and discussed previously at a conference (Sparks 1997).} with an image
stretch in the sense that white represents areas that were brighter in May and
black represents areas that were fainter. The reflected light variations due to
the Cepheid are clearly visible, as a series of bright and dark elliptical
rings, most prominently on the SW side. 

Although these images ($\rm FWHM\simeq1\farcs5$) do not have seeing as excellent
as the K08 frames, they reveal some important large-scale morphological
properties. First, the direct images show that the nebula is elongated in the
NE-SW direction, along a PA of about $35^\circ$. Second, the surface brightness
is generally higher on the SW side of the star than it is on the NE\null. This
is the appearance we would expect if the nebula has an overall cylindrical (or
bipolar) structure, tilted such that the SW lobe is in front of the plane of the
sky (giving it a relatively high surface brightness because of the smaller
scattering angle), and the NE lobe is behind the plane of the sky (large
scattering angle and lower surface brightness).\footnote{Note that an inclined
bipolar dust nebula illuminated by a central star will have a fundamentally
different appearance to an external observer than a bipolar planetary nebula
(PN)\null. Ionized gas (if optically thin) radiates isotropically, so that both
lobes in a PN will have the same brightness, whereas light scattered off dust
radiates preferentially in the forward direction, making the forward lobe
brighter than the one in the background.}

The elliptical rings seen in the ratio image in Fig.~5 strongly support this
interpretation. First, as noted in \S3, light variations will be averaged out if
there is an appreciable depth along the line of sight. Thus, the prominent
variations on the SW side indicate that there is a limited $z$-direction
extent of the nebula, i.e., it is not spherical. Moreover, on the SW side, the
arcs have the widest spacing, indicating a relatively wide spacing between the
paraboloids in the $x$ direction. As Fig.~1 shows, this wide spacing occurs in
the foreground. At least five separate rings can be seen on this side of the
star, again showing that this material must lie well in front of the plane of
the sky. By contrast, the rings are more closely spaced, and fade away rapidly,
on the NE side. This strongly suggests that the material here is in the
background, where the paraboloids are much more closely spaced in the $z$
direction, causing the light variations to average out.

The first, second, and third bright rings in the ratio image, where their
spacing is largest on the SW side, lie at radii of $\sim$$18''$, $34''$, and
$46''$. This suggests that the number of cycles in the lags is increasing in
steps of 1 at each of those radii [e.g., if the first frame had been taken at a
time of maximum of the Cepheid, the values of $(N_i+\Delta\phi_i)$ could be
0.44, 1.44, and 2.44 at these rings; or, if the first ring is too close to the
star to be resolved, the values could be 1.44, 2.44, and 3.44].

To give an indication of the orientation of the nebula suggested by these
features, we repeated the analysis of the previous subsection, again adopting
$\gamma=125^\circ$. For the 11 knots lying in the third quadrant ($180^\circ <
\beta < 270^\circ$), guided by the ring radii given above but with some degree
of arbitrariness, we set $N_i=1$ if $10\farcs96 < \theta_i < 17''$, $N_i=2$ if
$17'' < \theta_i < 30''$, and $N_i=3$ if $\theta_i > 30''$. These choices place
those knots well in front of the plane of the sky, as expected.  We then
artibrarily adjusted the $N_i$ values for the remainder of the 31 features so
that they generally fall along the plane established by the third-quadrant
knots. The resulting spatial distribution is plotted in the bottom panel of
Fig.~4. This rendition is consistent with all of the expectations, i.e., most of
the variable knots found by K08 lie in the foreground, most of them lie on the
SW side of the star (positive values of $d_x$) where the nebula is brightest,
and the knots in the SW quadrant have low values of $N_i$ ranging from 1 to 3.
For this set of $N_i$ values, the axis of the bipolar structure is tilted by
about $40^\circ$ with respect to the plane of the sky; but we emphasize that the
$N_i$ values for most of the knots were chosen arbitrarily to yield a flattened
distribution. The variable knots in the background are, actually, more probably
those on the front surface of the bipolar structure rather than along its axis
as shown.


\section{Polarimetry to the Rescue}

We have shown that the precise distance determination of K08 is invalidated by
the unjustified assumption that the variable dust knots in the \RS\ nebula lie
in the plane of the sky. In fact, we demonstrated that the nebula is plausibly
an elongated struture, highly tilted with respect to the plane of the sky.

Fortunately, the use of polarimetric imaging, at \HST\/ resolution, has the
potential to yield a geometric distance, much as it did for V838~Mon (see S08).
The degree of polarization of scattered light, $p$, depends strongly on the
scattering angle $\xi$, according to the classical Rayleigh function, $p=p_{\rm
max} \, (1-\cos^2\xi)/(1+\cos^2\xi)$, which has its maximum value at
$\xi=90^\circ$.  In the case of the V838~Mon echo, S08 indeed found a high
degree of linear polarization, $p_{\rm max}\simeq50\%$, at the $90^\circ$
scattering angle, from which they obtained a geometric distance to V838~Mon. As
described above (\S2), this distance has been verified independently through a
distance determination based on main-sequence fitting.

We thus have a simple recipe for determining a true geometric distance to \RS:

\begin{enumerate}

\item Obtain imaging polarimetry of the nebula over a 41.4-day pulsation cycle,
using the polarimetric mode of \HST's Advanced Camera for Surveys (ACS;
scheduled to be restored to service during the upcoming Servicing Mission~4).
Search the images for small knots that (a)~vary with the full light amplitude of
the Cepheid variation (showing that they are indeed isolated along the line of
sight), (b)~have the highest degree of linear polarization (demonstrating that
they lie near the plane of the sky), and (c)~have a constant degree of
polarization over the pulsation cycle (showing again that the line of sight is
undiluted by contributions from out of the plane of the sky).  If our
conclusions in \S5.2 regarding the morphology and viewing angle of the nebula
are correct, the best place to look for knots lying close to the plane of the
sky would be the NW and SE quadrants, or very close to the star in the NE and SW
quadrants. 

\item Determine the phase lag, $\Delta\phi_i$, for each such knot, relative to
that of \RS. Assume an integer cycle  count, $N_i$, based on an approximate
distance to the star, e.g., that given by Feast (2008). Fig.~1 shows that, for
the small values of $x$ observable with \HST, there will be no ambiguity in
$N_i$ when we know that the knot is near the plane of the sky, since we
certainly already know the distance to within a factor of 2. Then the distance
to the star is given by the equation in \S4.1.

\item Average the $d_i$ values for all detected highly polarized knots in order
to obtain the best mean distance value. Note that the equation for $p$ is
symmetric around the $90^\circ$ scattering angle so biasing toward the
foreground tends to be alleviated.

\end{enumerate}

Our proposed method does require that there exist isolated bright knots within
the nebula, in sufficient numbers that enough of them will in fact lie close to
the plane of the sky. The case of V838~Mon is encouraging, because \HST\/
resolves its light echo into a wealth of bright features down to the limit of
ACS resolution. Since the ground-based images of the \RS\ nebula already suggest
numerous fine structures, there are grounds for optimism. In addition to the
spatial resolution, ACS imaging provides the capability to go in very close to
the star, where, as Fig.~1 shows, there is a very large difference in scattering
angle with cycle count (along the $z$ direction), giving us a further tool to
resolve the count ambiguity.

\section{Conclusions}

\begin{enumerate}

\item Kervella et al.\ (2008; K08) have reported determination of an extremely
precise distance to the long-period Cepheid \RS,  based on measurements of phase
lags in the light echoes in its surrounding nebula.

\item Unfortunately, the K08 method was based on the assumption that the
light-variable nebular knots lie in the plane of the sky. We show that the
angular dependence of the scattering function, which strongly favors forward
scattering, produces a  a strong biasing toward the illuminated knots lying in
front of the plane of the sky. This bias invalidates the assumption made by K08.

\item In addition, we show that the statistical significance of K08's distance
measurement is low.

\item Feast (2008) has derived a highly flattened spatial distribution of the
variable knots based on K08's phase-lag data and a distance adopted from the
Cepheid period-luminosity relation, but we show that the derived distribution
depends entirely upon the arbitrarily adopted integer phase lags.

\item Based on two direct images and their ratio image, we show that it is
plausible that the \RS\ nebula has a cylindrical or bipolar structure tilted by
about $40^\circ$ with respect to the plane of the sky, but our derived spatial
distribution is not unique.

\item We argue that polarimetric imaging, at \HST\/ resolution, may potentially
yield a true geometric distance, if there is a sufficient number of isolated
small-scale knots in the nebula to identify the subset whose high linear
polarization shows them to lie very close to the plane of the sky.

\end{enumerate}

\begin{acknowledgements}

This work was partially supported by STScI grant GO-11217. We thank Michael
Feast, Bob Havlen, and Pierre Kervella and his team for valuable discussions.

\end{acknowledgements}

\begin{figure}
\centering
\includegraphics[width=0.5\textwidth]{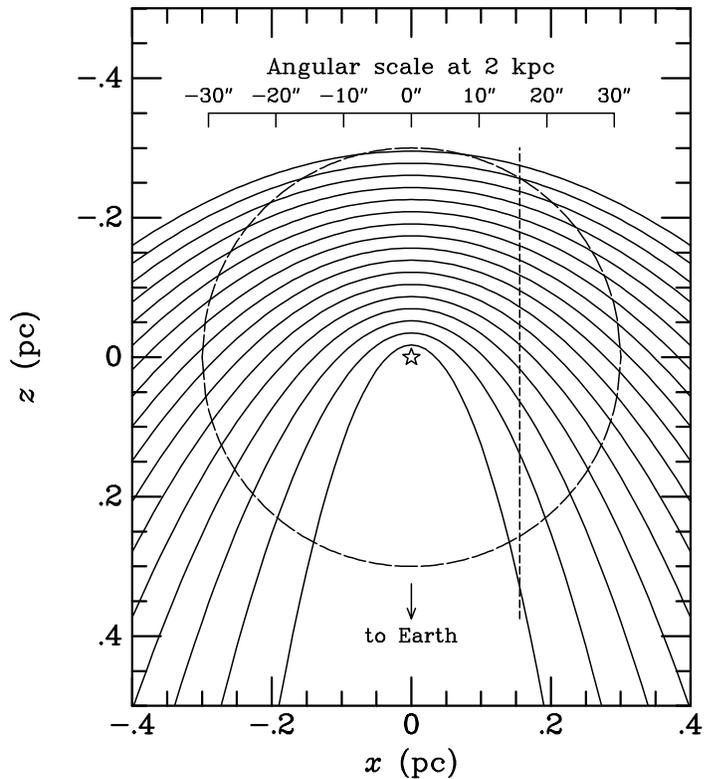} 

\caption{This figure illustrates the propagation of a train of nested light-echo
paraboloids into the dust surrounding \RS, as viewed from the Earth. Each
parabola marks a maximum in the 41.4-day variation of the illuminating star. The
$x$ and $z$ scales are in pc. The corresponding angular scale depends on the
distance to the star, and is shown at the top of the figure for a nominal
distance of 2~kpc. The dashed circle of radius 0.3~pc encloses the brightest
parts of the nebula, but fainter portions are seen out to at least twice that
radius. The vertical dashed line marks the location of Knot~5 at a projected
separation of $16\farcs0$. Knot~5 has a phase lag of 0.5~cycle relative to \RS,
so it must lie halfway between a pair of parabolas, but we do not know {\it
which\/} pair (see text).  }

\end{figure}


\begin{figure}
\centering
\includegraphics[width=0.2\textwidth]{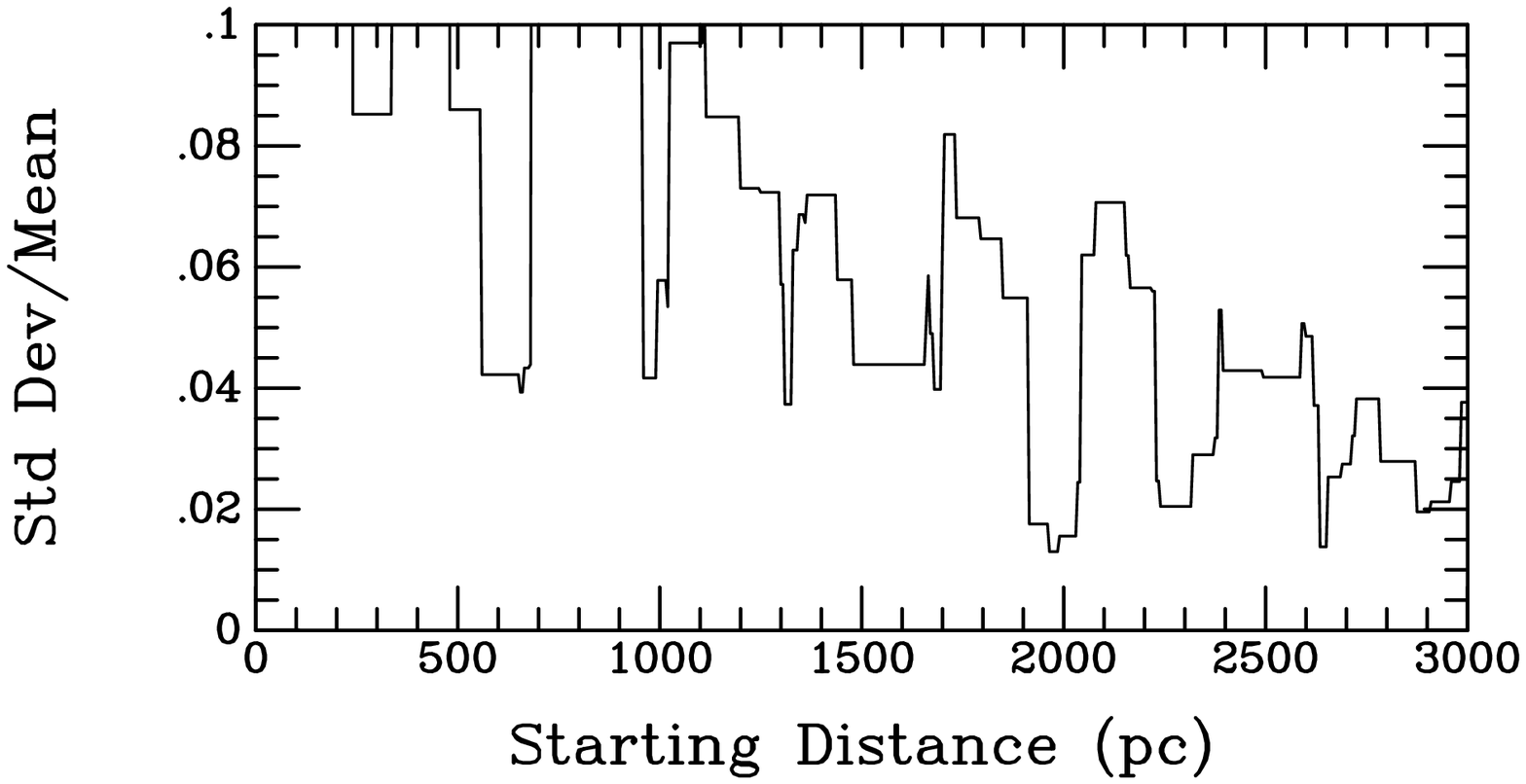}
\hskip0.20in
\includegraphics[width=0.2\textwidth]{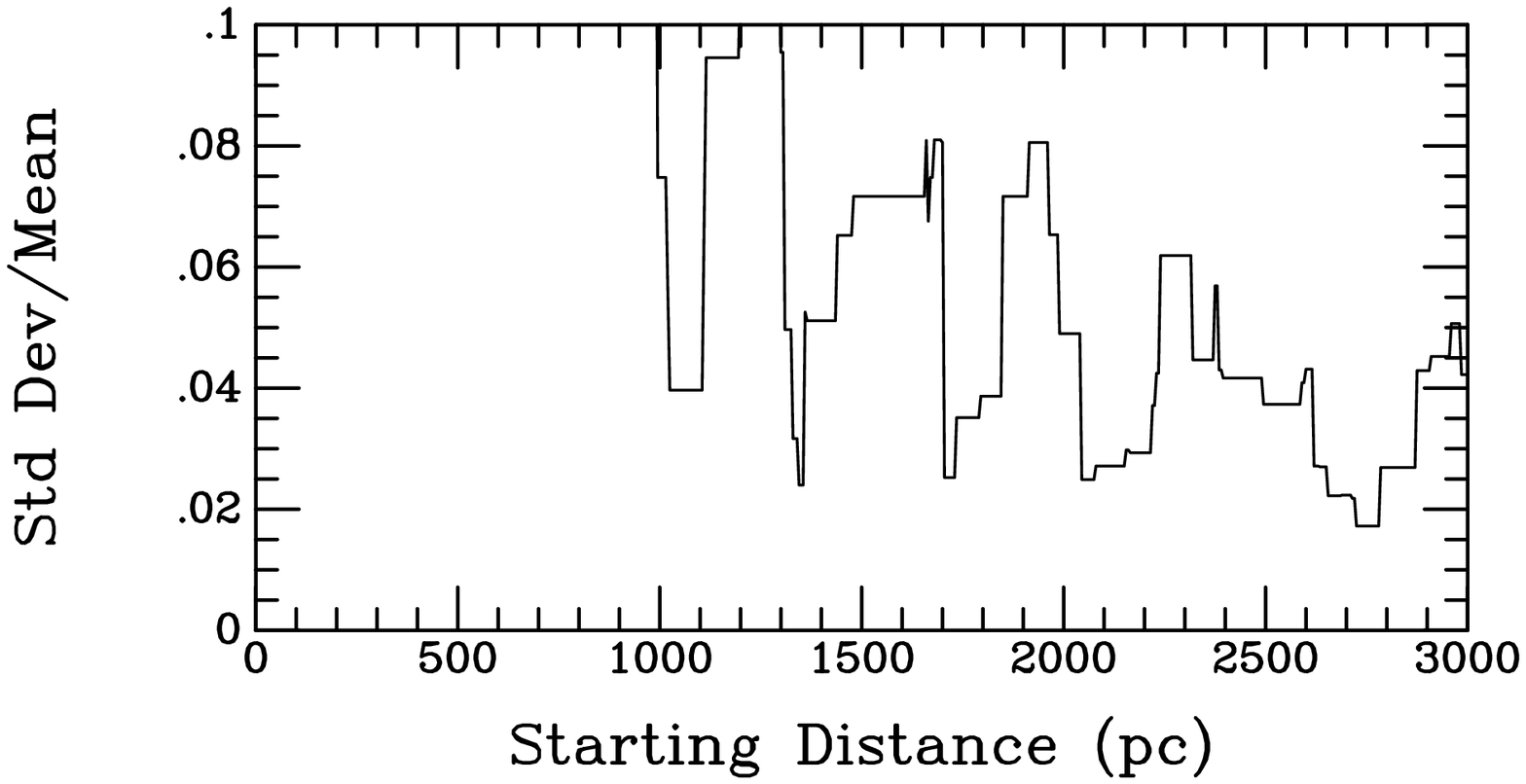}
\vskip0.20in
\includegraphics[width=0.2\textwidth]{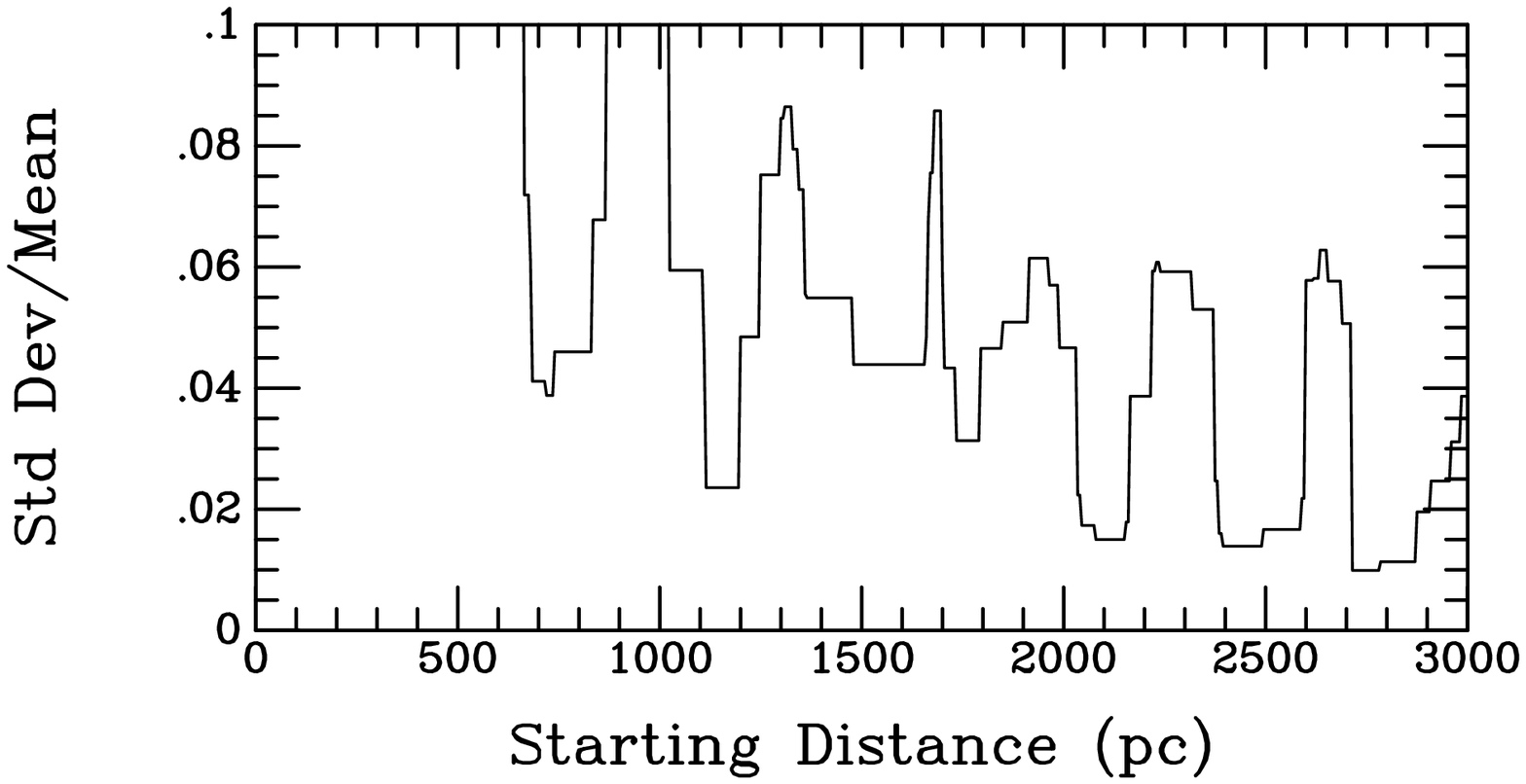}
\hskip0.20in
\includegraphics[width=0.2\textwidth]{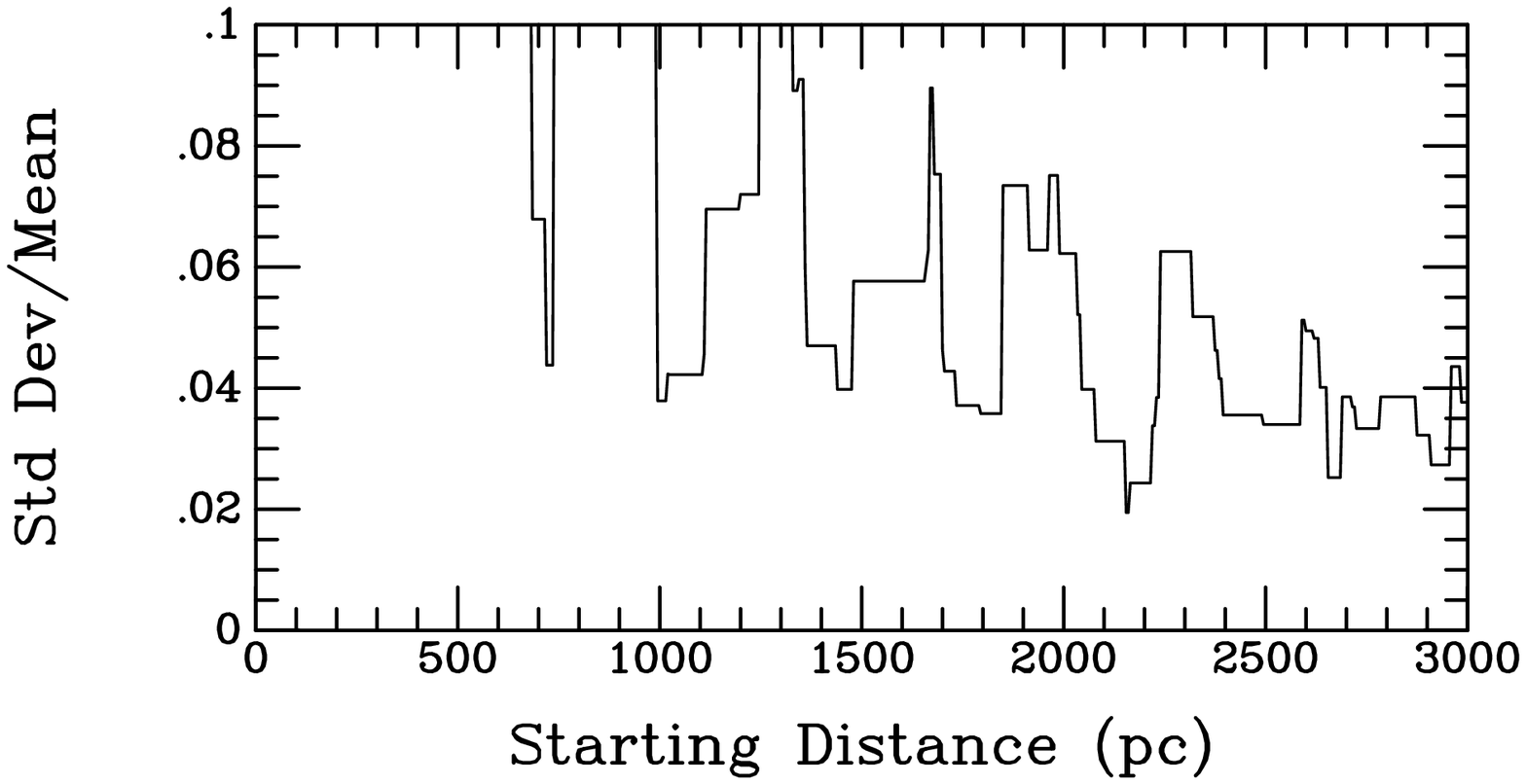}
\vskip0.20in
\includegraphics[width=0.2\textwidth]{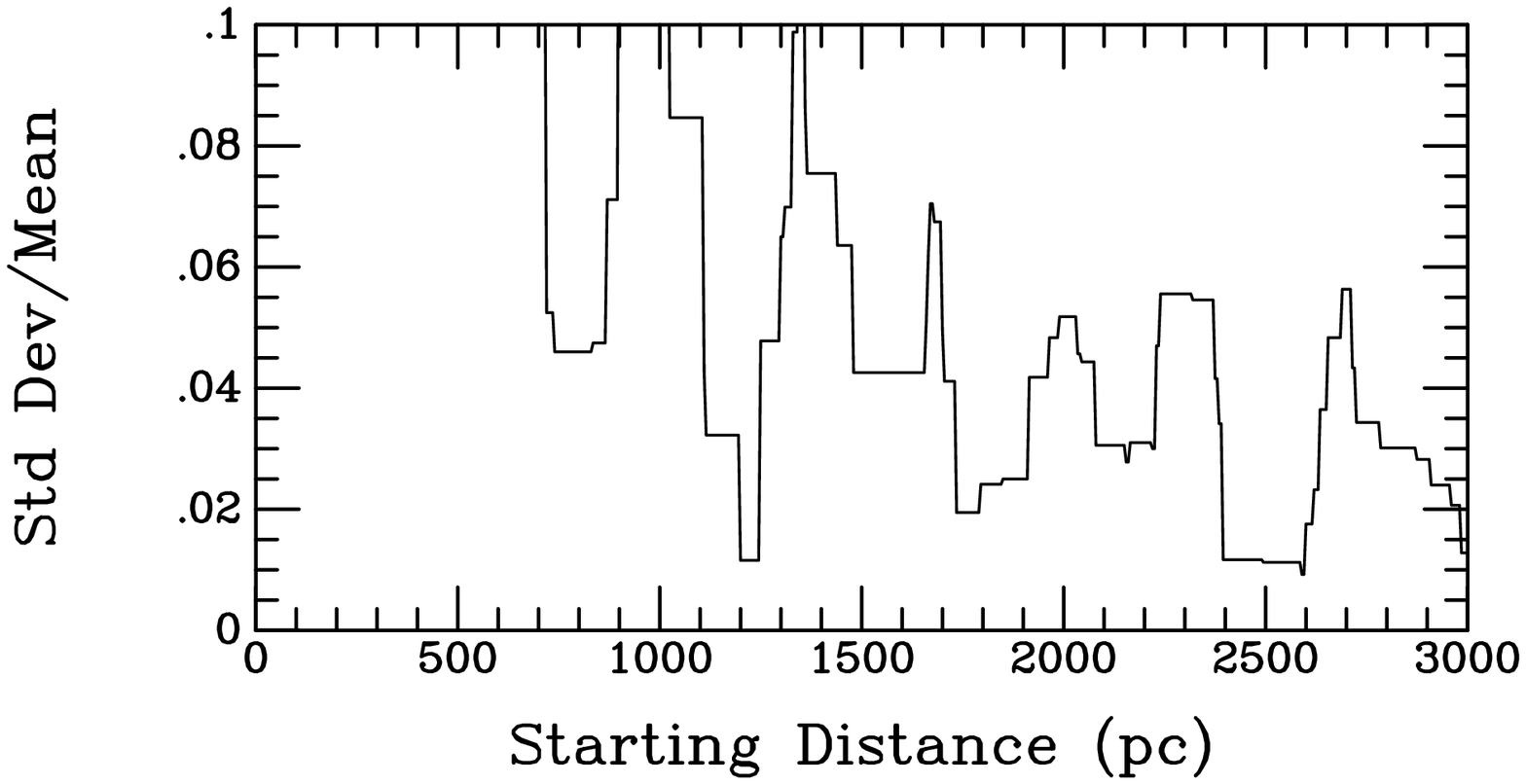}
\hskip0.20in
\includegraphics[width=0.2\textwidth]{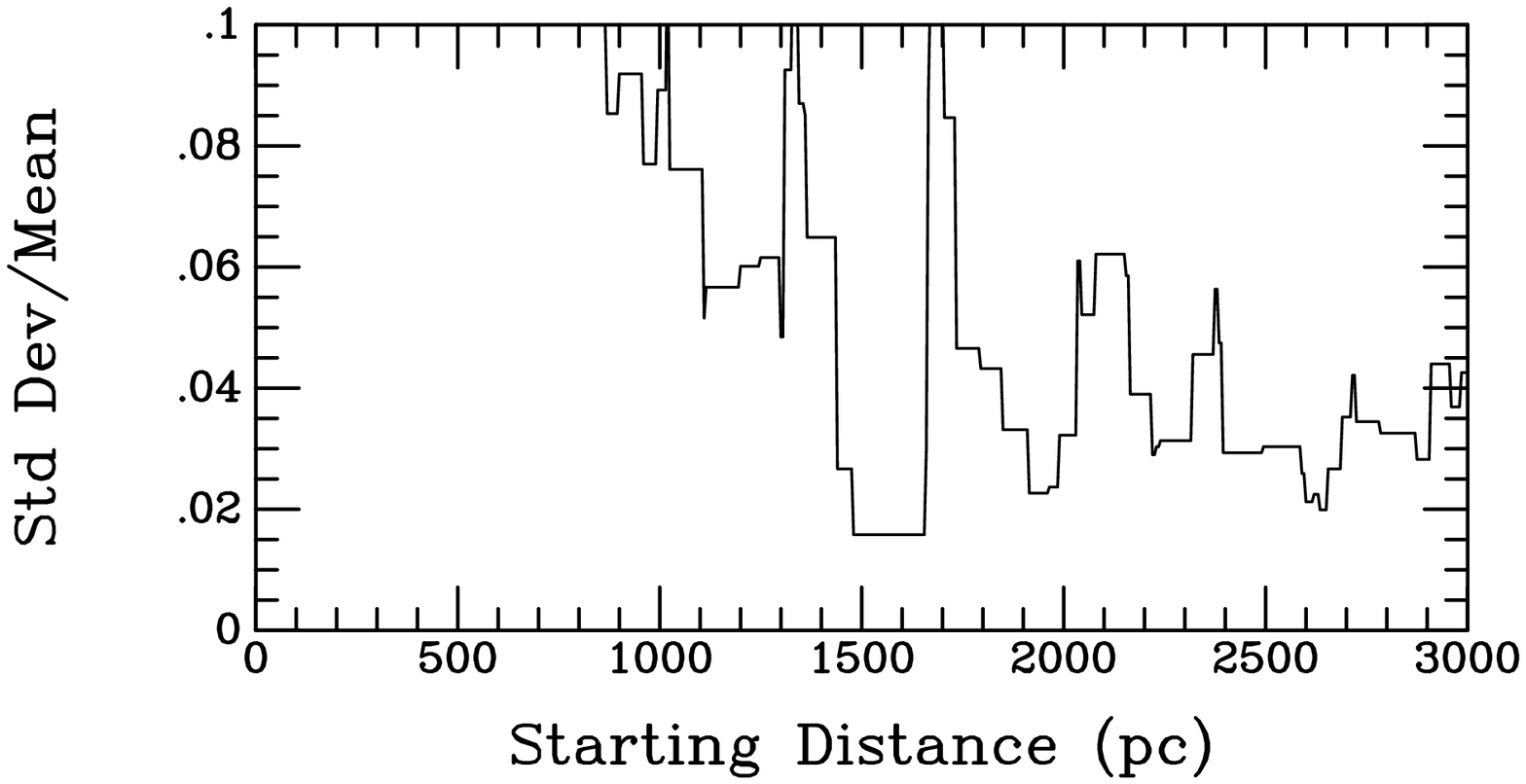}
\vskip0.20in
\includegraphics[width=0.2\textwidth]{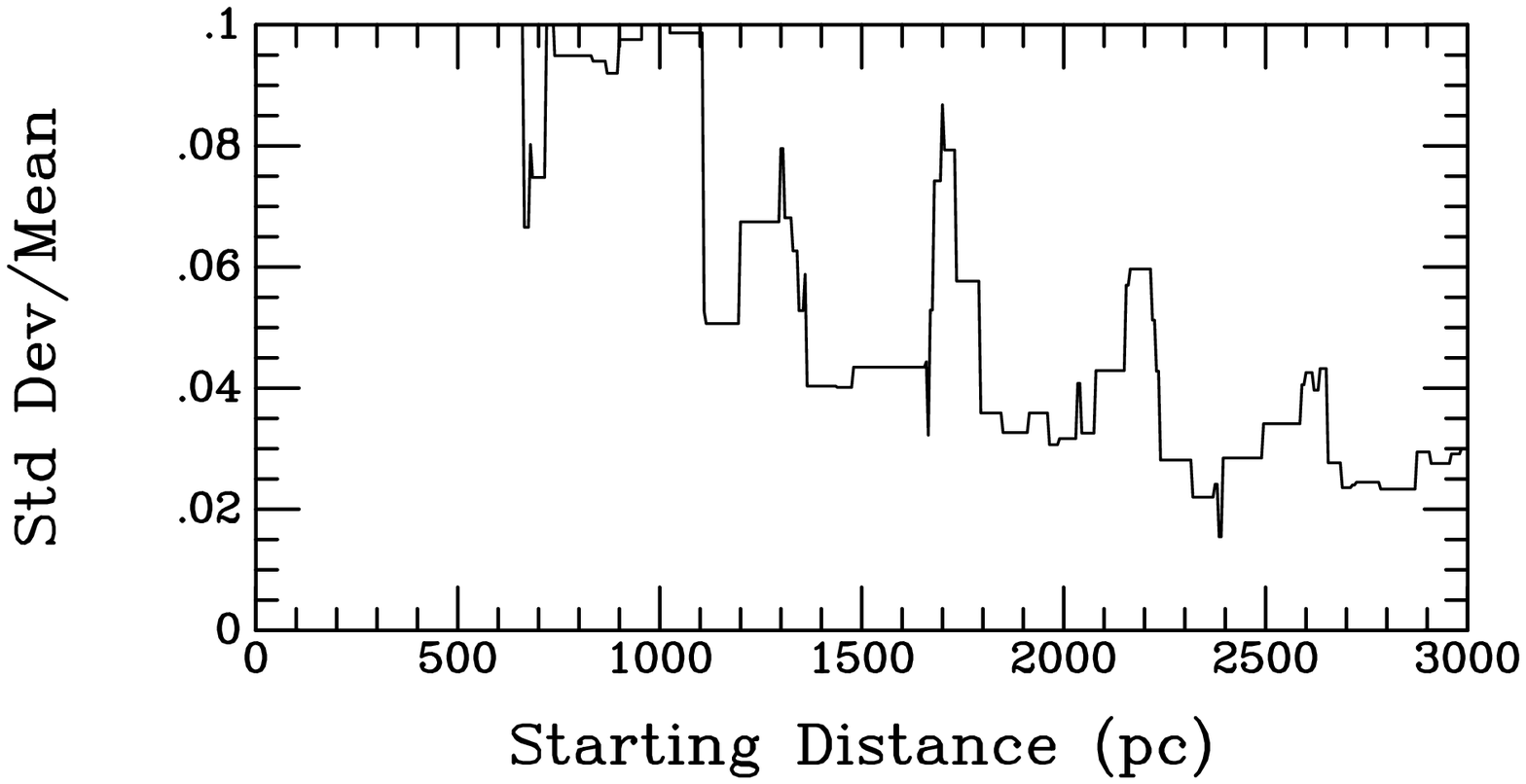}
\hskip0.20in
\includegraphics[width=0.2\textwidth]{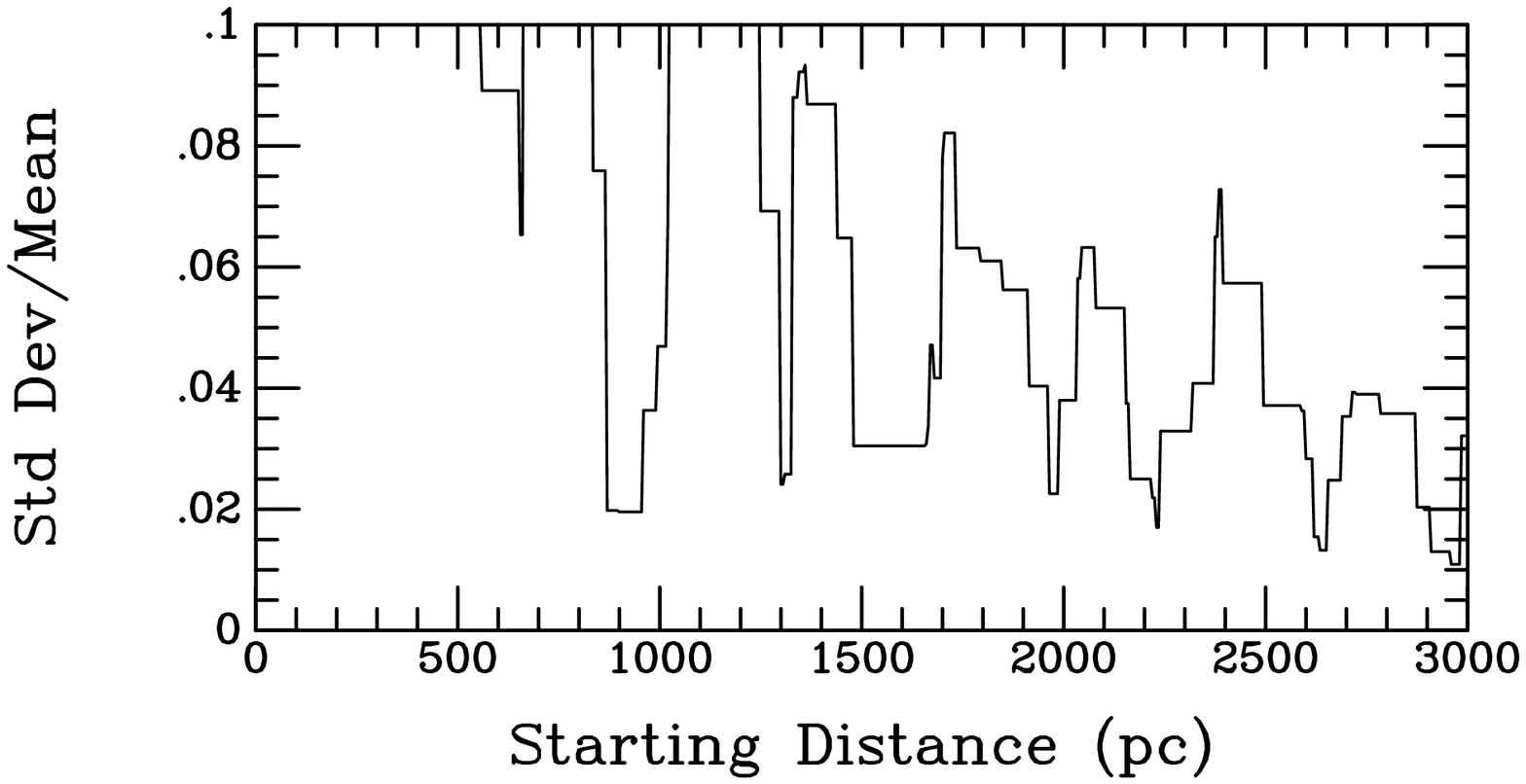}
\vskip0.20in
\includegraphics[width=0.2\textwidth]{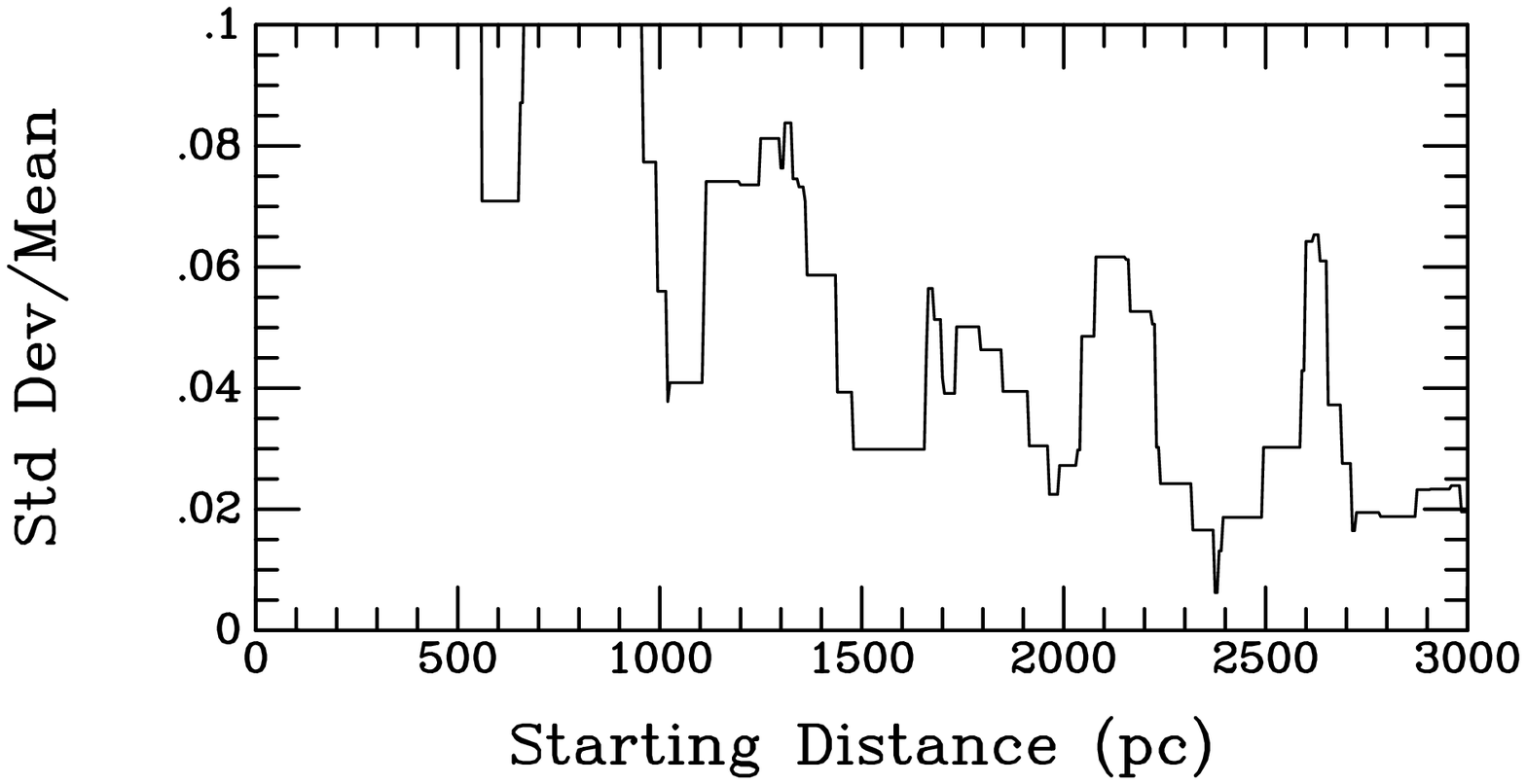}
\hskip0.20in
\includegraphics[width=0.2\textwidth]{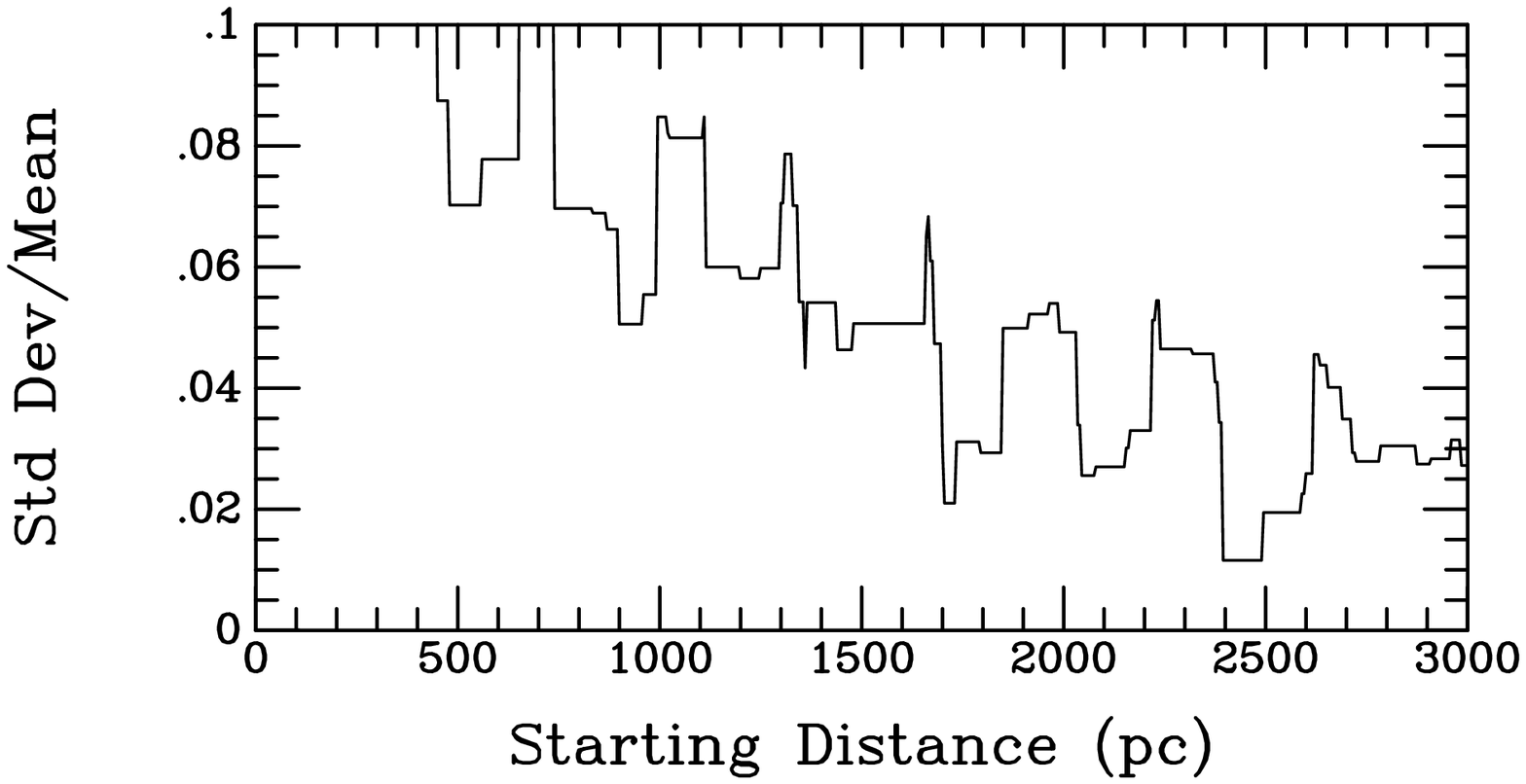}

\caption{These plots illustrate the method used by K08 to determine the distance
to \RS\ using the phase lags of individual variable knots in the surrounding
nebula. Over a range of assumed distances, plotted on the $x$ axis, the distance
$d_i$ to each knot is calculated from the equation in \S4.1, and the standard
deviation of the distances (after deleting the 2 highest and 2 lowest values),
divided by the mean distance, is plotted on the $y$ axis. The upper left panel
shows the result for the 10 high-weight knots detected by K08, and shows a
minimum relative standard deviation at a distance of 1992~pc. The remaining 9
panels show the result of the same calculation, using the same set of angular
separations, $\theta_i$, but with phase lags, $\Delta\phi_i$, produced by a
random-number generator. Two-thirds of these simulations produced minima lower
than that from the actual data. }

\end{figure}


\begin{figure}
\centering
\includegraphics[width=0.5\textwidth]{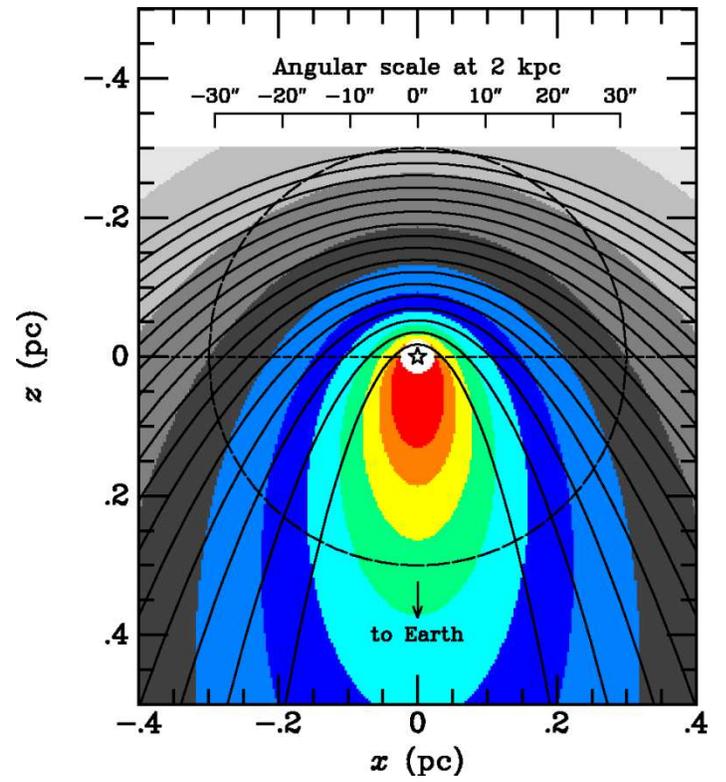} 

\caption{This figure repeats Fig.~1, with overplotted contours indicating the
value at each $(x,z)$ point of the scattering function, $\Phi(\xi)$, divided
by $r^2$, the square of the distance from the star. This ratio, multiplied
by the local dust density, is proportional to the contribution to the surface
brightness at a given line of sight along the $z$-direction as seen from Earth.
Each contour represents a step of a factor of 2 in $\Phi(\xi)/r^2$. The
strong biasing toward the brightest nebular knots lying in the foreground is
illustrated. The horizontal line at $z=0$ marks the plane of the sky.}

\end{figure}

\begin{figure}
\centering
\includegraphics[width=0.4\textwidth]{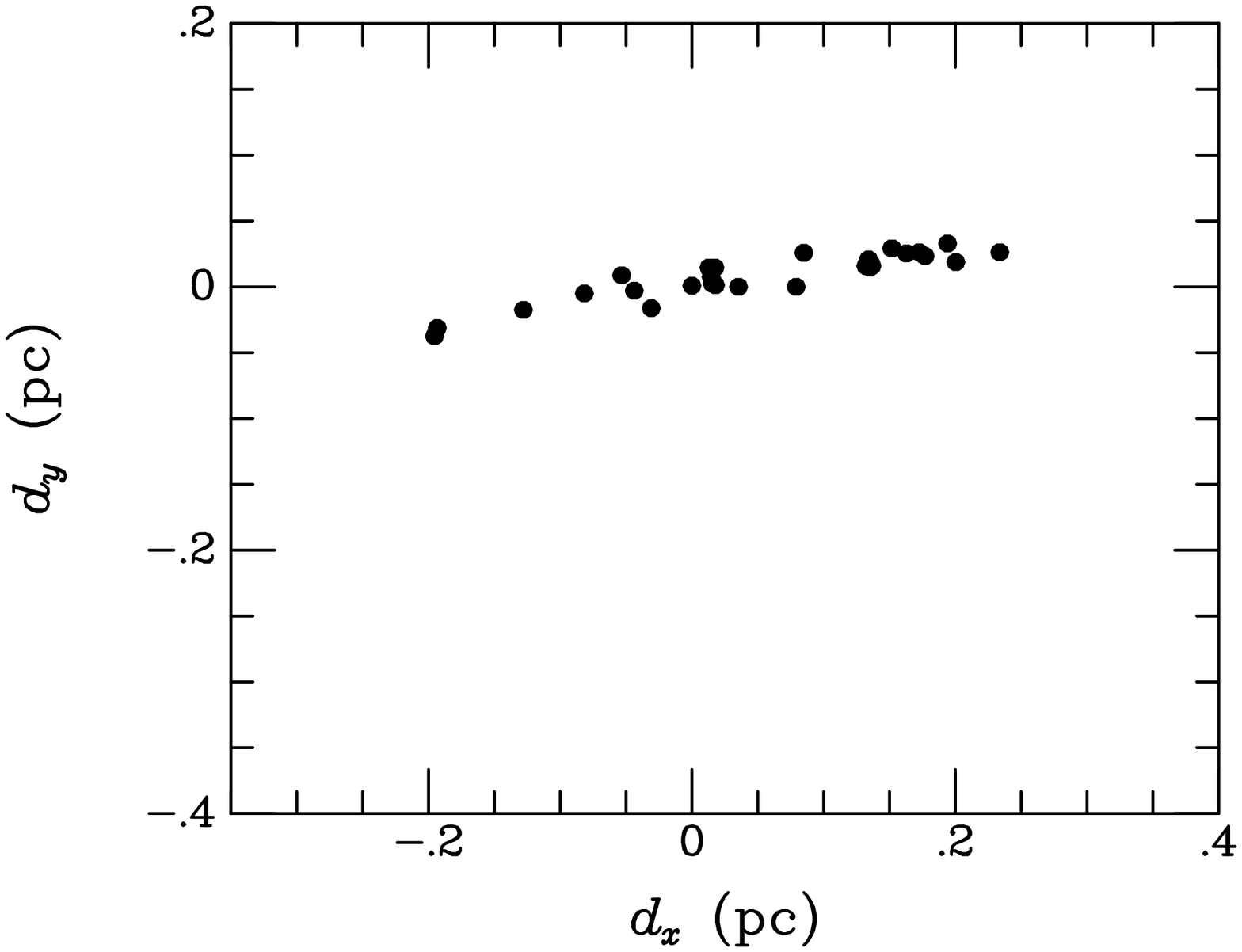}
\vskip 0.25in
\includegraphics[width=0.4\textwidth]{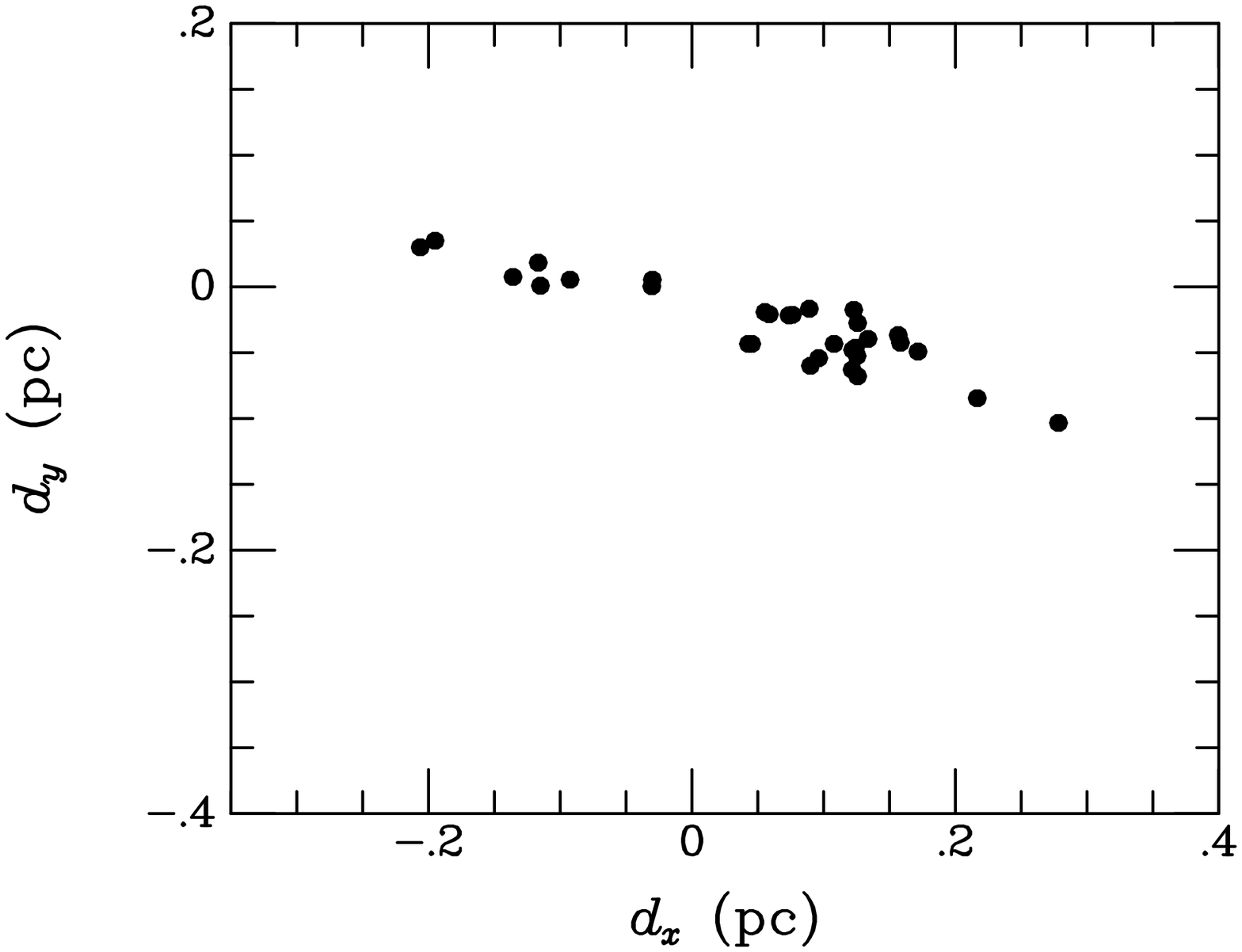}
\vskip 0.25in
\includegraphics[width=0.4\textwidth]{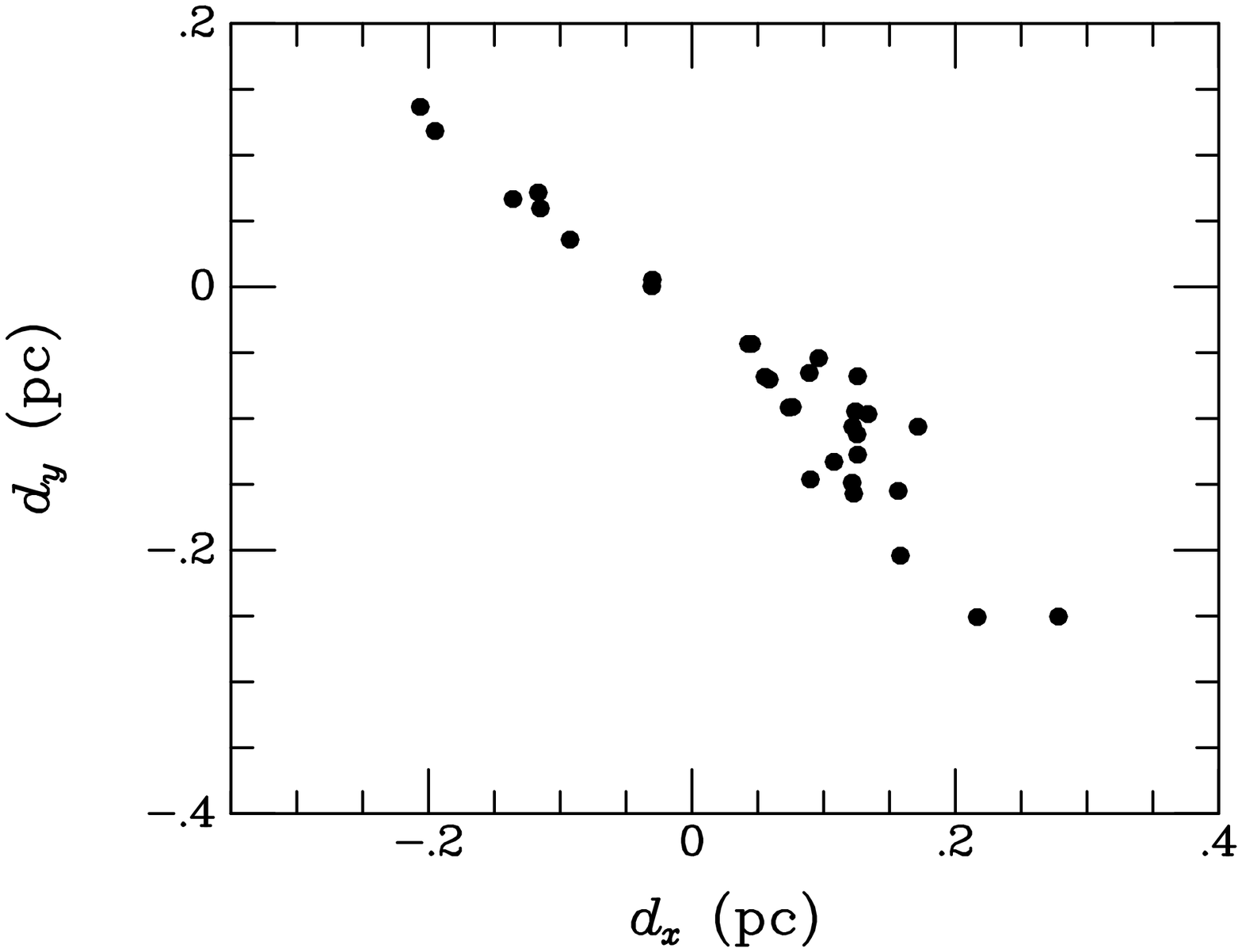}

\caption{{\bf Top panel:} Spatial distribution of 31 light-variable knots in the
\RS\ nebula, according to the set of $N_i$ values adopted by Feast (2008). The
distance from the plane of the sky is $d_y$ (equivalent to our $-z$), and $d_x$
is the distance in the plane of the sky perpendicular to a line of azimuth
$\gamma=80^\circ$. {\bf Middle panel:} Another possible spatial distribution,
with $\gamma=125^\circ$ and a different arbitrary set of $N_i$ values. {\bf
Bottom panel:} A third spatial distribution, derived as discussed in \S5.2, this
time setting the $N_i$ values in the third quadrant based on the ratio
image in Fig.~5, and then adjusting the remaining values to give a flattened
distribution. The Appendix lists the adopted values of $N_i$ for the three
distributions.}

\end{figure}

\clearpage


\begin{figure}
\centering
\includegraphics[width=0.25\textwidth]{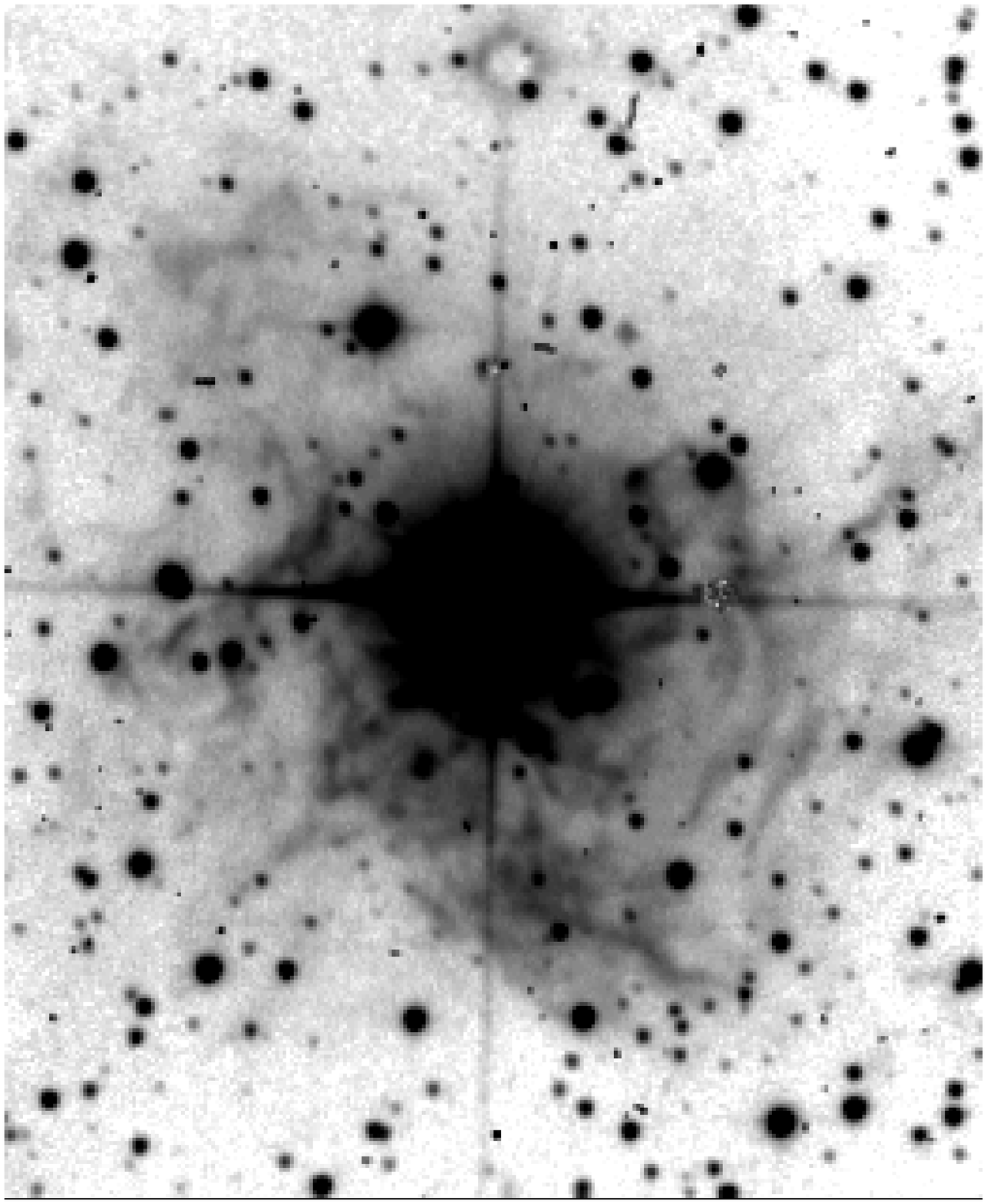}
\includegraphics[width=0.25\textwidth]{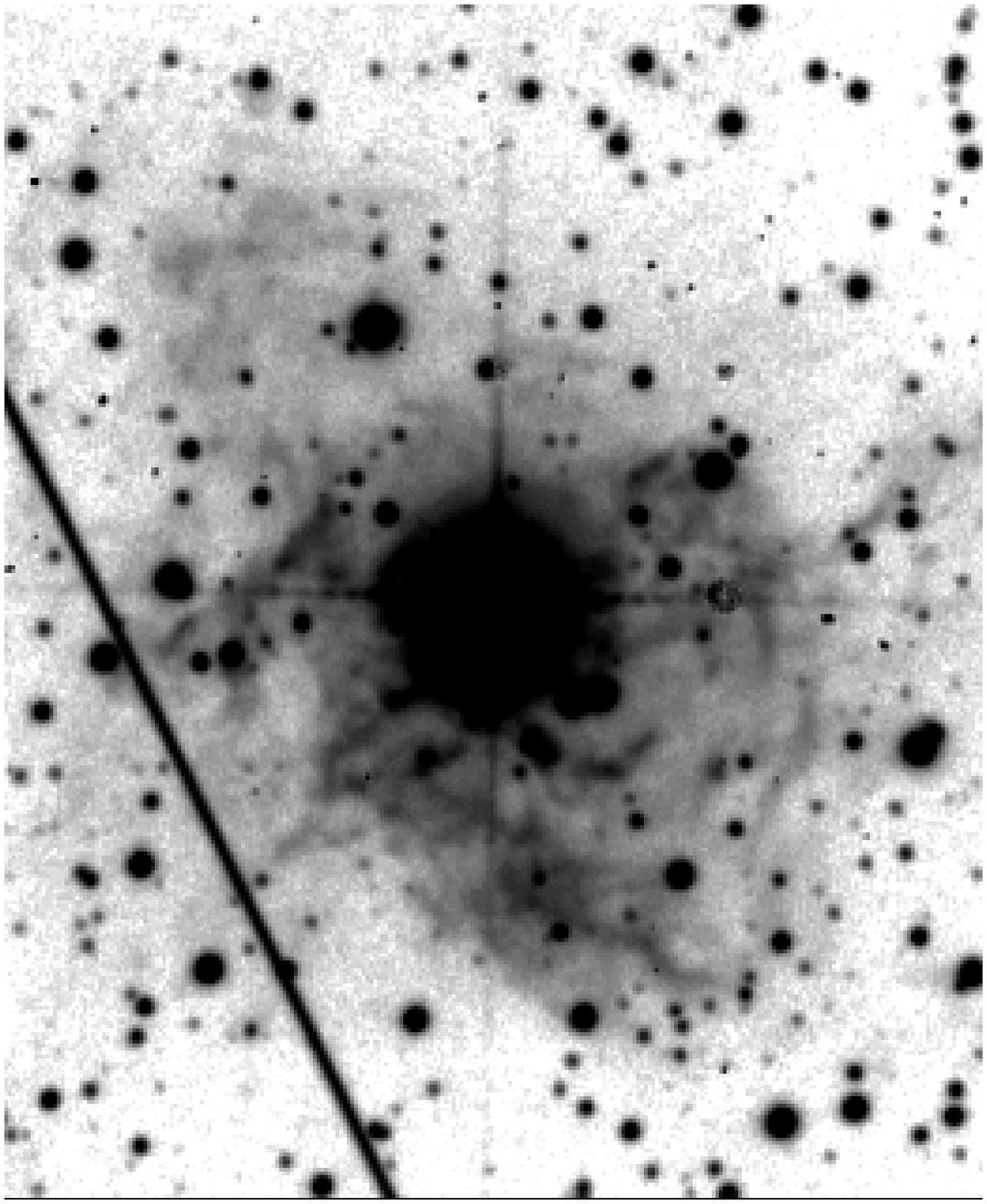}
\includegraphics[width=0.25\textwidth]{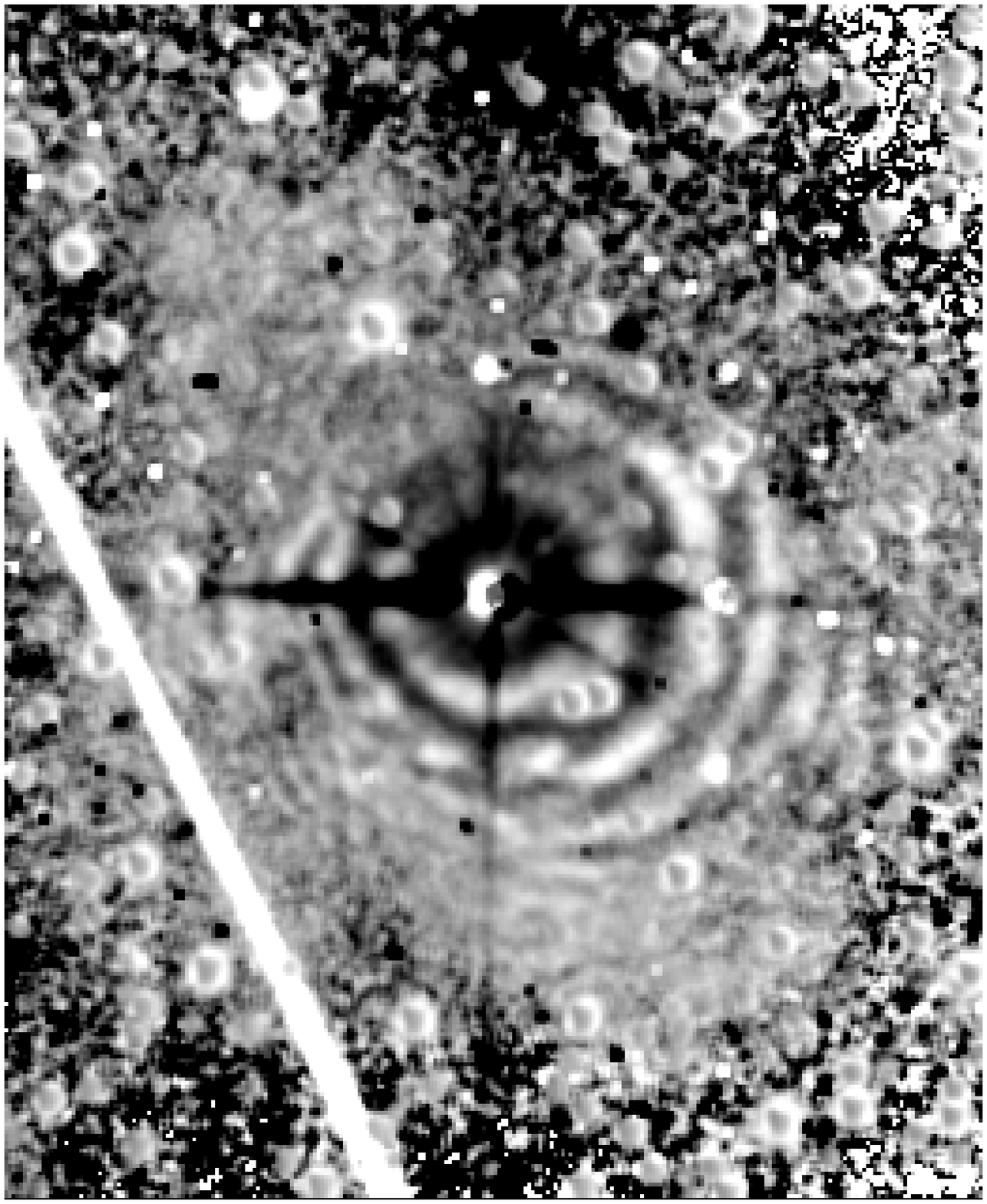}

\caption{The first two panels are ESO 3.6-m CCD images of \RS\ obtained on 1995
March~5 and May~4, a time separation of 1.44 Cepheid cycles. The third panel
shows the ratio of the two images. Stretches for the two direct images are
logarithmic, and linear for the ratio image. In the ratio image, white marks
regions brighter in May than March. North is at the top, east is on the left,
and the frames are $160''$ wide. There is a bright satellite streak in the May
image.  }

\end{figure}

\begin{appendix} 

\centerline{Appendix}

Table A.1 presents the values of $N_i$ used to derive the spatial distributions
shown in Fig.~4. The values of $\theta_i$, $\beta_i$, and $\Delta\phi_i$ have
been tabulated by Feast (2008).  Col.~1 is a running number, col.~2 gives the
values adopted by Feast, col.~3 is our arbitrary set that yields the flattened
distribution in the middle panel of Fig.~4, and col.~4 gives the set that yields
the more highly tilted spatial distribution in the bottom panel of Fig.~4.

\begin{table}
\caption{Values of $N_i$ Producing the Spatial Distributions in Figure 4}             
\label{table:1}      
\centering                          
\begin{tabular}{r r r r}        
\hline\hline                 
$i$ & $N_i$ & $N_i$ & $N_i$ \\
\hline                        
1   &  5  & 3 &	 3  \\
2   &  5  & 3 &  2  \\
3   &  5  & 3 &  2  \\
4   &  4  & 2 &  1  \\
5   &  4  & 2 &  1  \\
6   &  3  & 2 &  1  \\
7   &  3  & 2 &  1  \\
8   &  6  & 7 &  7  \\
9   &  4  & 4 &  6  \\
10  &  4  & 3 &  2  \\
11  &  3  & 2 &  1  \\
12  &  3  & 1 &  1  \\
13  &  4  & 3 &  2  \\
14  &  3  & 4 &  6  \\
15  &  4  & 3 &  2  \\
16  &  4  & 2 &  2  \\
17  &  4  & 4 &  5  \\
18  &  4  & 3 &  2  \\
19  &  5  & 3 &  2  \\
20  &  5  & 3 &  3  \\
21  &  5  & 4 &  2  \\
22  &  6  & 6 &  8  \\
23  &  7  & 4 &  2  \\
24  &  7  & 6 &  4  \\
25  &  7  & 5 &  4  \\
26  &  6  & 7 &  7  \\
27  &  7  & 8 &  12 \\
28  &  8  & 5 &  3  \\
29  &  8  & 9 &  12 \\
30  &  8  & 7 &  4  \\
31  &  8  & 8 &  5  \\
\hline                                   
\end{tabular}
\end{table}

\end{appendix}

\end{document}